# Spectral Analysis of Inhomogeneities Shows that the Elastic Stiffness of Random Composites Decreases with Increasing Heterogeneity

**Ehsan Ban**[1,2]
Department of Mechanical, Aerospace, and Nuclear Engineering, Rensselaer Polytechnic Institute, Jonsson Engineering Center, Room 2049, 110 8th Street, Troy, NY 12180;

[1] Corresponding author, e-mail: eban@seas.upenn.edu.
[2] Currently at University of Pennsylvania.

**Abstract**

Investigation of inhomogeneities has wide applications in different areas of mechanics including the study of composite materials. Here, we analytically study an arbitrarily-shaped isotropic inhomogeneity embedded in a finite-sized heterogeneous medium. By modal decomposition of the influence of the inhomogeneity on the deformation of the composite, an exact relation is presented that determines the variation of effective elastic stiffness caused by the presence of the inhomogeneity. This relation indicates that the effective elastic stiffness of a composite is always a concave function of the elastic modulus of the inhomogeneity, embedded inside the composite. Therefore, as the heterogeneity of elastic random composites increases, the rate of increase in effective stiffness caused by the stiffer constituents is smaller than the rate of its decrease due to the softer constitutions. So, weakly heterogeneous random composites become softer and less conductive with increasing heterogeneity at the same mean of constituent properties. We numerically evaluated the effective properties of about ten thousand composites to empirically support these results, characterize the nonaffinity of the displacement field in random composites, characterize the sample to sample variability of the effective properties and extend the results to the thermal conductivity of composites.

**1. Introduction**

The study of inhomogeneities has wide applications in mechanics. For instance, in determination of the effects of healthy and cancerous cells on the deformation of soft tissues (Schwarz and Safran, 2013; Mills et al., 2014; Shin et al., 2016) as well as predicting the properties of rubber in the presence of filler particles (Smallwood, 1944; Lopez-Pamies et al., 2013). Such composite materials are ubiquitously found in nature, whereas they can also be engineered to produce materials with desired properties. Stiff, lightweight composites are used as structural materials in cars and airplanes (Matthews and Rawlings, 1999);

composites with enhanced thermal conductivity are used to cool electronic devices (Chung, 2010).

Composites with given volume fractions of constituents exhibit a range of effective properties depending on the spatial arrangement of the constituent phases. In elasticity, if the strain field is uniform over the composite, the effective elastic stiffness equals the weighted arithmetic mean of the constituent moduli. This case represents the upper bound of the effective properties corresponding to the respective volume fraction of constituents. The lower bound is represented by the harmonic weighted average of the constituent properties (Hill, 1963; Milton, 2002). For any composition, specific geometric arrangements of phases can be found that match the properties corresponding to these bounds, indicating that these bounds are optimal (Christensen, 2005).

The effective properties of dilute mixtures can be approximated using the solution for the case of a single inhomogeneity embedded in a homogeneous medium. The inhomogeneity problems have been studied in various transport phenomena including elasticity (Smallwood, 1944; Eshelby, 1957; Zhou et al., 2013) and electromagnetism (Maxwell, 1881). Inhomogeneity problems have also been investigated in coupled transport phenomena (Ru, 2000) as well as in materials with nonlinear constitutive behaviors (Suquet, 1997). At larger volume fractions of inhomogeneities, however, the interactions between inhomogeneities causes the dilute approximations to be inaccurate. This difficulty has been addressed using various approximation methods (Christensen, 2005; Nemat-Nasser and Hori, 2013; Willis, 1981). Exact solutions exist for special cases such as for composites with specific constituent geometries and composites with specific distributions of constituents (Milton, 2002; Torquato, 1991). The estimates for random composites can be improved by accounting for the statistical moments of the spatial distribution of constituents, represented by multipoint correlation functions (Torquato, 2005, 1991). Similar approaches have been taken in studies of weakly heterogeneous composites where the differences between the constituent properties are small. Such methods have been utilized to study the bulk modulus of weakly heterogeneous random composites (Molyneux, J and Beran, M, 1965), evaluate the effective thermal conductivity of polycrystals using weak contrast expansions (Avellaneda and Bruno, 1990), and provide solutions for the effective stiffness and toughness of biomimetic random composites (Dimas et al., 2015a, 2015b). Similar approaches have been proposed to evaluate the effective nonlinear properties of composites (Castañeda, 1991).

In this paper, we first present the analytical spectral decomposition of the influence of an arbitrarily-shaped inhomogeneity on the effective elastic stiffness of a finite-sized elastic medium. The modal relation is illustrated and validated using the example of a spherical inhomogeneity embedded in a volumetrically expanding composite. Next, we used the analytical inhomogeneity results together with a weak contrast series expansion to study the effective properties of random elastic and conductive composites. Finally, numerical calculations are presented that empirically support the theoretical results, characterize the displacement of the random composites and its variability, and extend the presented results to the study of the thermal conductivity of composites.

## 2. Theory

This section describes an exact relation for the variation of the effective elastic stiffness, $C_{\text{eff}}$, of a finite-sized heterogeneous material caused by the addition of an arbitrarily-shaped isotropic inhomogeneity using the modal decomposition of the influence of the inhomogeneity. The inhomogeneity relation is then validated and illustrated using the example case of a spherical inhomogeneity embedded in a volumetrically expanding composite. The modal relation is used together with a weak contrast expansion to study the change in effective properties of random composites at various degrees of the microstructural heterogeneity of the composite.

### 2.1. Uniaxial stretch tests of a heterogeneous sample including an isotropic inhomogeneity

Uniaxial tests are considered to quantify the elastic stiffness of heterogeneous samples. A cubic heterogeneous elastic sample of dimensions $L_0$ and volume $V = L_0^3$ is considered. The effective elastic stiffness is measured by applying a small constant displacement $d_0$ in the direction $x_1$ to one of the cube face, $S_R$, while holding the opposite face fixed in $x_1$. Few other degrees of freedom are constrained to eliminate rigid body motions. The application of the boundary conditions produces a distribution of boundary traction[3] $\boldsymbol{w} = \{w_1, w_2, w_3\}$. The $x_1$ component of the traction $\boldsymbol{w}.\boldsymbol{e}_1 = w_1$ can be integrated over $S_R$ to evaluate the total boundary reaction[4]. Here, $e_1 = \{1,0,0\}$ denotes the unit vector in the direction $x_1$. The effective small-strain elastic stiffness resisting uniaxial deformation in the 1 direction, can be evaluated as $C_{\text{eff}} = \int_{S_R} w \cdot e_1 dS /(A_0 \varepsilon_0) = \int_{S_R} d_0 w \cdot e_1 dS /(V \varepsilon_0^2)$, where $\varepsilon_0 = d_0/L_0$ is the applied normal strain in the $x_1$ direction, and $A_0$ is the area of $S_R$ at rest. We denote this state as the reference state, state 0, with the boundary traction $w^{(0)}$ and effective elastic stiffness $C_{\text{eff}}^{(0)}$. In this state, the deformation of the sample is described by the displacement field, $u^{(0)}$ (Fig. 1a). Throughout the text, parenthesized superscripts refer to the state at which the deformation $d_0$ is applied. State 0 (Fig. 1a) refers to the state where the inhomogeneity is absent. At state 1 (Fig. 1b-d) an inhomogeneity with elastic modulus $E_{\text{inh}}^{(1)}$ is present, and at state 2 the elastic modulus of the inhomogeneity is $E_{\text{inh}}^{(2)}$.

---

[3] Throughout the article bold italic symbols refer to vector fields.

---

[4] For the two vectors $a = \{a_1, a_2, a_3\}$ and $b = \{b_1, b_2, b_3\}$, the componentwise dot product $a \cdot b = a_1 b_1 + a_2 b_2 + a_3 b_3$.

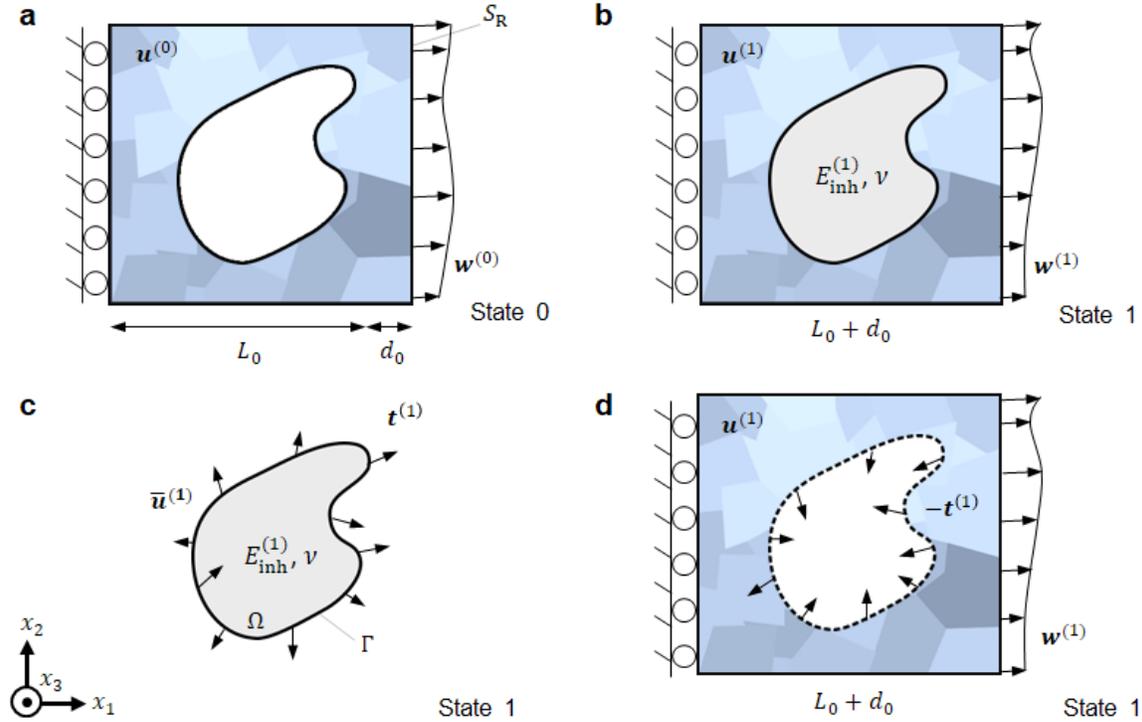

**Fig. 1.** (Color online) Schematic of the operation used to evaluate the variation in effective elastic stiffness caused by the presence of an arbitrarily-shaped inhomogeneity (central gray shape). (a) The reference state, state 0, where the inhomogeneity is absent. Effective elastic stiffness $C_{\text{eff}}$ is measured in a uniaxial tension test by displacing the right sample boundary in the $x_1$ direction by $d_0$ (b) State 1, where the uniaxial extension test is performed while an inhomogeneity of elastic modulus $E_{\text{inh}}^{(1)}$ and Poisson's ratio $v$ is added. (c) Free body diagram of the inhomogeneity with elastic modulus $E_{\text{inh}}^{(1)}$ in state 1. The inhomogeneity is deformed by the traction force $t^{(1)}$. (d) The inhomogeneity is replaced by the reactions to the boundary traction that the inhomogeneity is experiencing, $-t^{(1)}$. The inhomogeneity is a Lipschitz domain $\Omega$ in $\mathbb{R}^3$ with smooth boundary $\Gamma$.

### 2.2. Spectral decomposition of the influence of an isotropic inhomogeneity on effective elastic stiffness

In this section, the influence of an inhomogeneity on effective elastic stiffness is evaluated using a modal decomposition and superposition operations. The effect of the inhomogeneity is replaced by the tractions that it exchanges with the matrix while under the influence of the displacements to the composite's boundary. The boundary tractions are decomposed into orthonormal components. Then, the change in boundary traction caused by the application of the force component corresponding to each eigenfunction is evaluated using the reciprocal theorem (Love, 2011). It is found that invariants exist for individual modes, which determine the change in sample deformation as a function of the elastic modulus of the inhomogeneity. The invariants are then used to evaluate the variation of effective elastic stiffness caused by the addition of the inhomogeneity. The result is an exact analytical relation between the variation of effective properties of a finite-sized initially heterogeneous medium and the elastic modulus of the inhomogeneity, showing that effective elastic stiffness is a concave function of the properties of an inhomogeneity embedded inside the composite. This result is used later to demonstrate that in random composites the mean effective elastic stiffness decreases with increasing the composite's heterogeneity.

We begin by considering the case where an inhomogeneity with the elastic modulus $E_{\text{inh}}^{(1)}$ is added to the heterogeneous cubic sample (Fig. 1b). If the boundary conditions described in state 0 (Fig. 1a) are applied to this sample, the boundary traction changes to $\boldsymbol{w}^{(1)}$. The effective elastic stiffness and the displacement field in this case are denoted by $C_{\text{eff}}^{(1)}$ and $\boldsymbol{u}^{(1)}$, respectively. We refer to this state of the system as state 1. If we consider the free body diagram of the inhomogeneity in state 1, its boundary is deformed by the displacement $\overline{\boldsymbol{u}}^{(1)} = \{\overline{u}_1^{(1)}, \overline{u}_2^{(1)}, \overline{u}_3^{(1)}\}$ and the traction force $\boldsymbol{t}^{(1)} = \{t_1^{(1)}, t_2^{(1)}, t_3^{(1)}\}$. As commonly practiced in small-strain elasticity, we replace the inhomogeneity by the forces that it applies to the surrounding medium, $-t^{(1)}$. Using the reciprocal theorem (Love, 2011) between the states 0 and 1 yields

$$\int_{S_R} d_0 w^{(1)} \cdot e_1 dS = \int_{S_R} d_0 w^{(0)} \cdot e_1 dS + \int_\Gamma t^{(1)} \cdot \bar{u}^{(0)} d\Gamma. \tag{1}$$

Therefore, the change in the effective elastic stiffness can be expressed as

$$C_{\text{eff}}^{(1)} = C_{\text{eff}}^{(0)} + \frac{1}{V\varepsilon_0^2} \int_\Gamma t^{(1)} \cdot \bar{u}^{(0)} d\Gamma. \tag{2}$$

In the next section, we use the modal decomposition of the force and displacement in the inhomogeneity to rewrite Eq. (2) in terms of $E_{\text{inh}}$.

### 2.2.1. Spectral decomposition of the traction and displacement at the boundary of the inhomogeneity

Spectral decompositions have been thoroughly studied in problems of elasticity (Ciarlet, 2013; Lebedev and Vorovich, 2003) and are commonly used in studying the normal modes of elastic vibrations (Thomson, 1996). While, eigenvectors form an orthonormal basis in the spectral form of stiffness matrices for frames and discrete structures (Bathe, 1982), eigenfunctions are present in the spectral analysis of continua. In this section, we use eigenfunctions to analyze the spectral decomposition of the displacement and traction at the boundary of an isotropic inhomogeneity in the absence of body forces (Fig. 1c).

We begin by considering the tractions and boundary displacement of an isotropic inhomogeneity with an elastic modulus $E_{\text{inh}} = 1$ and Poisson's ratio $\nu$. We denote the traction of the inhomogeneity with unit elastic modulus by $t'$. Following the work of Eshelby (Eshelby, 1959), the Green's function $G_{ij}$ can be used to find displacement at the boundary of the inhomogeneity due to the boundary traction as $\bar{u}_i = \int_\Gamma G_{ij} t'_j \, d\Gamma$. Here, the functions $t'$ and $\bar{u}$ are elements of a Hilbert space. The integral equation relating $t'$ and $\bar{u}$ can be represented by $\bar{u} = \mathcal{L} t'$ where $\mathcal{L}$ is a linear operator associated with the inner product $(t'^{(1)}, t'^{(2)}) = \int_\Gamma t'^{(1)} \cdot t'^{(2)} \, d\Gamma$ (Reed and Simon, 1981). $\mathcal{L}$ is positive because $(t', \mathcal{L} t') = \int_\Gamma \bar{u} \cdot t' \, d\Gamma \geq 0$, since $\int_\Gamma \bar{u} \cdot t' \, d\Gamma$ is twice the external work expended to deform the inhomogeneity by $\bar{u}$, which is stored as elastic strain energy and is non-negative. The reciprocal theorem of elasticity shows that $(\mathcal{L} t'^{(1)}, t'^{(2)}) = (t'^{(1)}, \mathcal{L} t'^{(2)})$, and therefore $\mathcal{L}$ is also self-adjoint. Furthermore, over the compact domain $\Omega$ the integral representation of $\mathcal{L}$ indicates that $\mathcal{L}$ is compact (Reed and Simon, 1981). Therefore, the spectral theorem of compact self-adjoint operators can be used to decompose the boundary traction and displacements as $t' = \sum_{j=1}^n \alpha_j \boldsymbol{\phi}_j$ and $\bar{u} = \sum_{j=1}^n \lambda_j \alpha_j \boldsymbol{\phi}_j$ (Reed and Simon, 1981). Here, $\lambda_j$ and $\boldsymbol{\phi}_j$ are eigenvalues and eigenfunctions, $\alpha_j = (t', \boldsymbol{\phi}_j)$, and $n$ is the number of eigenfunctions, which is not necessarily finite in this case. The basis set, $\boldsymbol{\phi}_j$ functions, can be chosen such that the eigenfunctions are orthonormal, i.e. $(\boldsymbol{\phi}_i, \boldsymbol{\phi}_j) = 0$ for $i \neq j$ and $(\boldsymbol{\phi}_i, \boldsymbol{\phi}_j) = 1$ for $i = j$. $\mathcal{L} \boldsymbol{\phi}_j = \lambda_j \boldsymbol{\phi}_j$, and therefore, the positive property of $\mathcal{L}$ indicates that $(\boldsymbol{\phi}_j, \mathcal{L} \boldsymbol{\phi}_j) = \lambda_j \geq 0$.

Next, we consider the change in the orthonormal decomposition when the inhomogeneity's elastic modulus is changed to $E_{\text{inh}}$ at the same Poisson's ratio. We recall that boundary traction is related to stress as $t = \sigma n$, where $\sigma$ is the stress tensor, and $n$ denotes the unit outward normal at the boundary. For an isotropic material, Hook's law can be used to express stress as $E_{\text{inh}} \left[ \frac{\nu}{(1+\nu)(1-2\nu)} \mathbf{I} \nabla \cdot \overline{\boldsymbol{u}} + \frac{1}{2(1+\nu)} (\nabla \overline{\boldsymbol{u}} + (\nabla \overline{\boldsymbol{u}})^{\text{T}}) \right]$ where $\mathbf{I}$ is the identity tensor.

Therefore, $\boldsymbol{t}$ is related to $\overline{\boldsymbol{u}}$ as $\boldsymbol{t} = E_{\text{inh}} \left[ \frac{\nu}{(1+\nu)(1-2\nu)} \mathbf{I} \nabla \cdot \overline{\boldsymbol{u}} + \frac{1}{2(1+\nu)} (\nabla \overline{\boldsymbol{u}} + (\nabla \overline{\boldsymbol{u}})^{\text{T}}) \right] \boldsymbol{n}$.

This relation also shows that if $E_{\text{inh}} = 1$ then $\boldsymbol{t}' = \left[ \frac{\nu}{(1+\nu)(1-2\nu)} \mathbf{I} \nabla \cdot \overline{\boldsymbol{u}} + \frac{1}{2(1+\nu)} (\nabla \overline{\boldsymbol{u}} + (\nabla \overline{\boldsymbol{u}})^{\text{T}}) \right] \boldsymbol{n}$. These equations indicate that any traction $\boldsymbol{t}$ and boundary displacement $\overline{\boldsymbol{u}}$ for an inhomogeneity with modulus $E_{\text{inh}}$ can be expressed as $\boldsymbol{t} = E_{\text{inh}} \boldsymbol{t}'$ and $\overline{\boldsymbol{u}}$, where $\boldsymbol{t}'$ and $\overline{\boldsymbol{u}}$ correspond to an inhomogeneity with $E_{\text{inh}} = 1$ and the same $\nu$. Therefore, using the spectral decomposition of $\boldsymbol{t}'$ and $\overline{\boldsymbol{u}}$, for an inhomogeneity with modulus $E_{\text{inh}}$, $\boldsymbol{t}$ and $\overline{\boldsymbol{u}}$ can be expressed as $\boldsymbol{t} = E_{\text{inh}} \boldsymbol{t}' = E_{\text{inh}} \sum_{i=1}^{n} \alpha_i \boldsymbol{\phi}_i$ and $\overline{\boldsymbol{u}} = \sum_{i=1}^{n} \lambda_i \alpha_i \boldsymbol{\phi}_i$.

**2.2.2. Influence of a single mode on effective properties**

In this Section, we use Eq. (2) to find $C_{\text{eff}}^{(1)}$ at state 1. If we consider that only the reaction to the traction related to the $j$'th eigenfunction, $-\boldsymbol{t}_j^{(1)} = -E_{\text{inh}}^{(1)} \alpha_j^{(1)} \boldsymbol{\phi}_j$, is present in the stretch tests, then we can rewrite Eq. (1) as

$$\int_{S_R} d_0 \boldsymbol{w}_j^{(1)} \cdot \boldsymbol{e}_1 dS = \int_{S_R} d_0 \boldsymbol{w}_j^{(0)} \cdot \boldsymbol{e}_1 dS + \int_{\Gamma} \boldsymbol{t}_j^{(1)} \cdot \overline{\boldsymbol{u}}^{(0)} d\Gamma, \tag{3}$$

where $\boldsymbol{w}_j^{(1)}$ is the traction at the boundary $S_R$ due to the displacement $d_0$ and traction $-\boldsymbol{t}_j^{(1)}$. The orthonormal property of the eigenmodes, $\boldsymbol{\phi}_j$, yields that $\int_{\Gamma} \boldsymbol{t}_j^{(1)} \cdot \overline{\boldsymbol{u}}^{(0)} d\Gamma = E_{\text{inh}}^{(1)} \alpha_j^{(1)} \alpha_j^{(0)} \lambda_j$. Therefore,

$$\int_{S_R} d_0 \boldsymbol{w}_j^{(1)} \cdot \boldsymbol{e}_1 dS = \int_{S_R} d_0 \boldsymbol{w}_j^{(0)} \cdot \boldsymbol{e}_1 dS + E_{\text{inh}}^{(1)} \alpha_j^{(1)} \alpha_j^{(0)} \lambda_j. \tag{4}$$

Now, we consider another state where the inhomogeneity has a modulus $E_{\text{inh}}^{(2)}$ at the same Poisson's ratio, and the reaction to the traction from the $j$'th mode is applied $-t_j^{(2)} = E_{\text{inh}}^{(2)} \alpha_j^{(2)} \phi_j$. We denote this state as state 2. If the reciprocal theorem is applied between states 1 and 2 when we apply the traction from the $j$'th mode, $t_j$, we obtain

$$\int_{S_R} d_0 w_j^{(1)} \cdot e_1 dS + \int_{\Gamma} t_j^{(2)} \cdot \bar{u}^{(1)} d\Gamma$$

$$= \int_{S_R} d_0 w_j^{(2)} \cdot e_1 dS + \int_{\Gamma} t_j^{(1)} \cdot \bar{u}^{(2)} d\Gamma. \tag{5}$$

By using the orthogonality of the modes, we can rewrite Eq. (5) as

$$\int_{S_R} d_0 w_j^{(1)} \cdot e_1 dS + E_{\text{inh}}^{(2)} \alpha_j^{(2)} \alpha_j^{(1)} \lambda_j$$

$$= \int_{S_R} d_0 w_j^{(2)} \cdot e_1 dS + E_{\text{inh}}^{(1)} \alpha_j^{(1)} \alpha_j^{(2)} \lambda_j. \tag{6}$$

If the inhomogeneity is not deformed in the $j$'th mode in state 0, increasing its stiffness does not alter the effective properties. However, if the element deforms, combining Eqs. (4 and 6) and the counterpart of Eq. (4) corresponding to states 2 and 0, gives

$$\frac{\left(\alpha_j^{(0)} - \alpha_j^{(1)}\right)}{E_{\text{inh}}^{(1)} \alpha_j^{(1)}} = \frac{\left(\alpha_j^{(0)} - \alpha_j^{(2)}\right)}{E_{\text{inh}}^{(2)} \alpha_j^{(2)}} = c_j. \tag{7}$$

Equation (7) indicates that the quantity $\left(\alpha_j^{(0)} - \alpha_j^{(i)}\right)/\left(E_{\text{inh}}^{(i)} \alpha_j^{(i)}\right)$ is invariant of $E_{\text{inh}}^{(i)}$. We denote this invariant by $c_j$. We note that $c_j = 0$ when the deformation of the inhomogeneity is prescribed by the boundary conditions and will not change by the application of $t_j^{(1)}$. It can be noted that $c_j$ is invariant of $E_{\text{inh}}^{(i)}$, whereas it depends on the prescribed boundary conditions. For a nonzero $c_j$, by replacing $\alpha_j^{(1)}$ in Eq. (4) and using the definition of $c_j$ from Eq. (7) we obtain

$$\int_{S_R} d_0 w_j^{(1)} \cdot e_1 dS = \int_{S_R} d_0 w_j^{(0)} \cdot e_1 dS + \frac{\lambda_j \left(\alpha_j^{(0)}\right)^2}{c_j + 1/E_{\text{inh}}^{(1)}}. \tag{8}$$

Equations (3 and 8) indicate that $\int_\Gamma t_j^{(1)} \cdot \bar{u}^{(0)} d\Gamma = \lambda_j \left(\alpha_j^{(0)}\right)^2 / (c_j + 1/E_{\text{inh}}^{(1)})$.

### 2.2.3. A modal relation for effective elastic stiffness

The integral $\int_\Gamma t^{(1)} \cdot \bar{u}^{(0)} d\Gamma$ can be decomposed as $\sum_{j=1}^n \int_\Gamma t_j^{(1)} \cdot \bar{u}^{(0)} d\Gamma$. Therefore, Eq. (2) can be rewritten as

$$C_{\text{eff}}^{(1)} = C_{\text{eff}}^{(0)} + \frac{1}{V \varepsilon_0^2} \sum_{j=1}^n \frac{\lambda_j \left(\alpha_j^{(0)}\right)^2}{c_j + 1/E_{\text{inh}}^{(1)}}, \tag{9}$$

where $C_{\text{eff}}^{(0)}$ and $C_{\text{eff}}^{(1)}$ are the effective stiffness in states 0 and 1. Equation (9) provides a relation between the change of the effective elastic stiffness of the composite and the variation of elastic modulus of an inhomogeneity. We note that $c_j$ is necessarily non-negative since otherwise, it is possible to find a variation of the modulus $E_{\text{inh}}^{(1)} = -1/c_j$, that leads to the divergence of $C_{\text{eff}}^{(1)}$, which is physically impossible.

Equation (9) indicates that effective elastic modulus is, in general, a concave function of the properties of the inhomogeneity:

$$\frac{\partial^2 C_{\text{eff}}}{\partial E_{\text{inh}}^2} \leq 0. \tag{10}$$

The exceptional case where equality holds is described in the Discussion Section.

The inhomogeneity relation of Eq. (9) is different from most previous results by the fact that we did not assume a specific shape for the inhomogeneity. Many previous exact theories hold for inhomogeneities with specific geometric shapes, for example, spheres, ellipsoids, or polygons. Furthermore, we did not assume a homogeneous medium for the sample containing the inhomogeneity. Our theoretical derivations hold even if the inhomogeneity is part of a finite-sized composite consisting of phases with highly different properties. This assumption distinguishes our derivations from the results obtained using the T-matrix approaches (Gubernatis and Krumhansl, 1975) which are based on the assumption of small fluctuations of the strain field. Additionally, we assumed a finite-sized domain for the composite. Finally, the condition of the continuity of the medium is not required for our analysis. The reciprocal theorem for frame structures (Maxwell, 1864) can be used to derive Eq. (9) in the case of discrete structures such as cellular solids, networks of fibers (Ban et al., 2016), and metamaterials. The spectral theorems linear algebra and the eigendecomposition of compliance matrices can be used in that case. Special cases of Eq. (9) for inhomogeneities with one and two modes of deformation such as elastic rods and Euler-Bernoulli beams were explored in networks of beams (Ban et al., 2016).

**2.3. The illustrative example of a spherical inhomogeneity in a volumetrically expanding composite**

To validate the modal approach and apply it to a specific case, we evaluated the change in the effective bulk modulus of a spherical composite that includes a spherical inhomogeneity. The inhomogeneity and the surrounding matrix have elastic moduli $E_{inh}$ and $E_m$ and Poisson's ratio $\nu$. The two spheres have radii $r_i$ and $r_o$ and are centered at the same point $O$ (Fig. 2). The effective bulk modulus was measured by displacing the outer surface of the composite, $S_o$, by a small displacement, $d_0$, in the radial direction, $e_r$. We then measured the resulting traction $w$ over $S_o$ and evaluated the effective bulk modulus as $K_{eff} = \int_{S_o} w \cdot e_r dS /(S_0 \delta_0) = \int_{S_o} d_0 w \cdot e_r dS /(V\delta_0^2)$. Here, the volumetric strain $\delta_0 = 3d_0/r_o$, and $S_0$ is the initial surface area of the spherical composite.

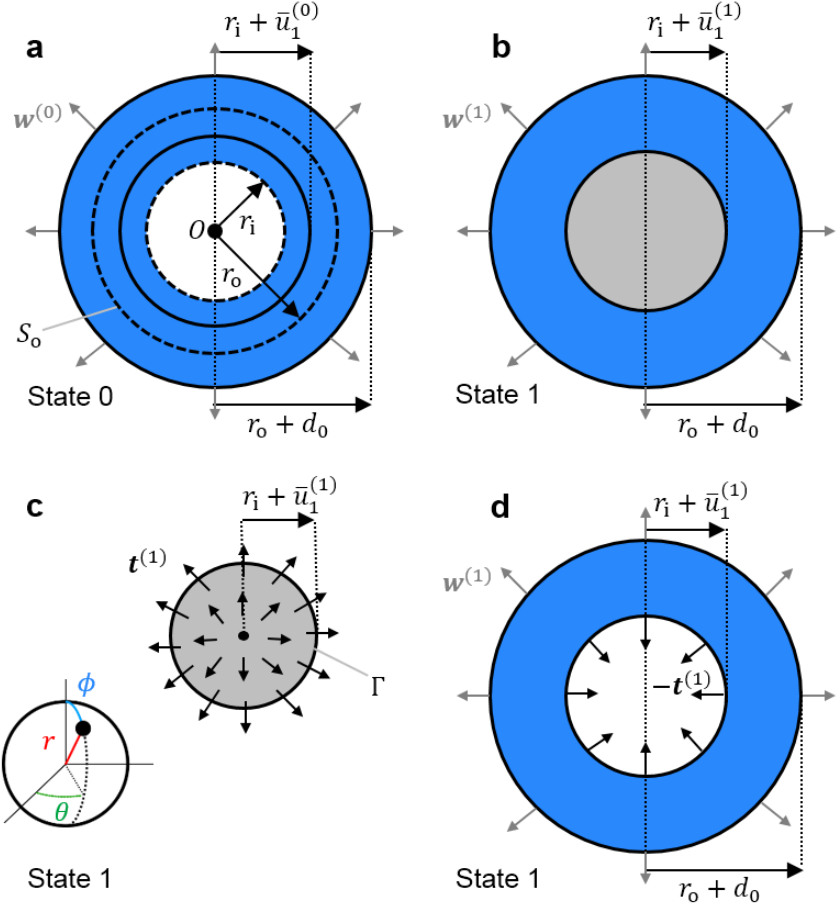

**Fig. 2.** (Color online) The illustrative example and the operations used to evaluate the variation of the effective bulk modulus of a spherical elastic composite caused by the presence of a spherical inhomogeneity (gray). (a) The reference state, state 0, where the inhomogeneity is absent. Effective bulk modulus $K_{\text{eff}}$ is measured in volumetric expansion tests by the radial displacement of the outer surface of the composite by $d_0$. The configurations before and after the application of deformation are plotted using dashed and solid lines, respectively. (b) State 1, where the volumetric expansion test is performed while a spherical inhomogeneity of elastic modulus $E_{\text{inh}}^{(1)}$ and Poisson's ratio $\nu$ is added to the composite. (c) Free body diagram of the inhomogeneity of elastic modulus $E_{\text{inh}}^{(1)}$ in state 1. The inhomogeneity is deformed by the traction force $t^{(1)}$. We denote the domain and boundary of the inhomogeneity by $\Omega$ and $\Gamma$. (d) The inhomogeneity is replaced by the reactions to the boundary traction that the inhomogeneity is experiencing, $-t^{(1)}$. A planar representation of the traction force is shown for simplicity.

Similar to the case of the arbitrarily-shaped inhomogeneity, we consider the two states 0 and 1 where the inhomogeneity is absent (Fig 2a) and where an inhomogeneity of elastic modulus $E_{inh}^{(1)}$ is present, respectively (Fig. 2b). The free-body diagram of the inhomogeneity shows that at state 1 it will have uniform surface traction and displacement of $t^{(1)}$ and $\bar{u}^{(1)}$, respectively (Fig. 2c). Without a loss of generality, we assume that the composite is fixed at the point $O$. We take out few additional degrees of freedom to ensure that the composite does not rotate about $O$. The deformation of the composite is, therefore, spherically symmetric, and the inhomogeneity only deforms in a single mode, volumetric expansion. The associated surface traction and displacement are $t^{(1)} = t_1^{(1)} e_r$ and $\bar{u}^{(1)} = \bar{u}_1^{(1)} e_r$. In this case, the modal decomposition, presented in Section 2.2.1, can be employed to express $t^{(1)}$ in terms of an eigenfunction, $t^{(1)} = E_{inh}^{(1)} \alpha_1^{(1)} \phi_1$, where $\phi_1$ is the only eigenfunction. The normal property of the modes requires that the inner product $(\phi_1, \phi_1) = 1$. Therefore, $\phi_1 = (1/2r_i\sqrt{\pi}) e_r$. This eigenfunction represents the volumetric expansion of the inhomogeneity, and the factor $(1/2r_i\sqrt{\pi}) e_r$ ensures that the inner product equals 1. The eigendecomposition also indicates that $\bar{u}^{(1)} = \lambda_1 \alpha_1^{(1)} \phi_1$, where $\lambda_1$ is the only present eigenvalue.

Next, the bulk modulus of the inhomogeneity, $K_{inh}^{(1)}$ can be used to relate the magnitude of the traction $t_1^{(1)}$ and volumetric strain $\delta_{inh}^{(1)}$ as $t_1^{(1)} = K_{inh}^{(1)} \delta_{inh}^{(1)}$. For an isotropic inhomogeneity $K_{inh}^{(1)} = E_{inh}^{(1)}/(3(1-2\nu))$. Therefore, if a displacement $\bar{u}_1^{(1)}$ is present at the boundary of the inhomogeneity, $t_1^{(1)} =$ $E\bar{u}_1^{(1)}/((1-2\nu)r_i)$. In this case, comparison with the modal form of $t$ indicates that for a displacement $\bar{u}_1^{(1)}$, $\alpha_1^{(1)} = 2\bar{u}_1^{(1)}\sqrt{\pi}/(1-2\nu)$. Noting that in this case, $\bar{\boldsymbol{u}}^{(1)} = \bar{u}_1^{(1)}\boldsymbol{e}_r = \lambda_1\alpha_1^{(1)}\boldsymbol{\phi}_1$ the eigenvalue can be evaluated as $\lambda_1 = r_i(1-2\nu)$.

Equation (8) can be in this case rewritten as

$$\int_{S_o} d_0\boldsymbol{w}_1^{(1)}\cdot\boldsymbol{e}_r dS = \int_{S_o} d_0\boldsymbol{w}_1^{(0)}\cdot\boldsymbol{e}_r dS + \frac{\lambda_1\left(\alpha_1^{(0)}\right)^2}{c_1 + 1/E_{\text{inh}}^{(1)}}, \tag{11}$$

where $\alpha_1^{(0)}$ corresponds to the deformation of the inhomogeneity in state 0.

$\alpha_1^{(0)}$ and the invariant $c_1$ can be evaluated for this problem using the solution for a pressurized hollow sphere (Bower, 2009). In general, the solution for the displacement in the radial direction takes the form $u(r) = Ar + B/r^2$ where $A$ and $B$ depend on the geometry, material properties, and boundary conditions. We begin by $\alpha_1^{(0)}$ in the absence of internal pressure, or traction in the inner surface of the hollow sphere. By prescribing the displacement $d_0$ at the outer boundary of the composite and considering the constitutive equation, we evaluate the constants $A$ and $B$. We then find the displacement at the inner surface of the sphere $\bar{u}_1^{(0)}$, and evaluate $\alpha_1^{(0)} = 2\bar{u}_1^{(0)}\sqrt{\pi}/(1-2\nu)$ as

$$\alpha_1^{(0)} = \frac{6r_i r_o^2\sqrt{\pi}d_0(1-\nu)}{(1-2\nu)\left(2r_o^3(1-2\nu) + r_i^3(1+\nu)\right)}. \tag{12}$$

Similarly, we prescribed a uniform radial traction force of magnitude $t_1^{(1)}$ at the interior surface of the hollow sphere and evaluated the displacement. We then used the expression $c_1 = (u_1^{(1)} - u_1^{(0)})/(\lambda_1 t_1^{(1)})$ and evaluated the invariant $c_1$ as

$$c_1 = \frac{(r_o^3 - r_i^3)(1+v)}{E_m\left(2r_o^3(1-2v) + r_i^3(1+v)\right)}.\tag{13}$$

Equations (11–13) and the relation between the boundary tractions and $K_{\text{eff}}$ can be used to evaluate the change in $K_{\text{eff}}$ by the addition of the inhomogeneity as

$$K_{\text{eff}}^{(1)} = K_{\text{eff}}^{(0)} + \frac{1}{V\delta_0^2}\frac{\lambda_1\left(\alpha_1^{(0)}\right)^2}{c_1 + 1/E_{\text{inh}}^{(1)}}.\tag{14}$$

Equation (14) demonstrates that $K_{\text{eff}}$ is a concave function of $E_{\text{inh}}^{(1)}$. Equation (14) was evaluated for inhomogeneities of different stiffnesses and radii (Fig. 3). The numerical values used to plot Eq. (14) are described in Section 3.2.

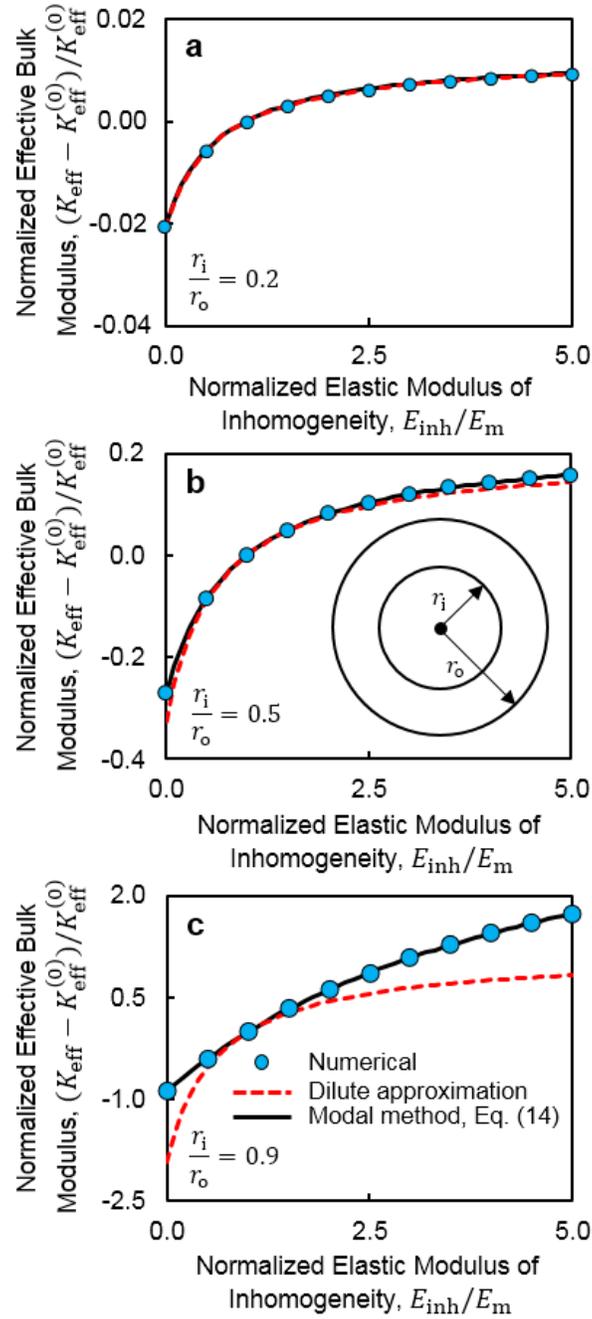

**Fig. 3.** (Color online) Effective bulk modulus of a spherical elastic composite containing a spherical inhomogeneity. In all cases, the effective bulk modulus was a concave function of the elastic modulus of the inhomogeneity. The elastic moduli of the inhomogeneity and the surrounding matrix are $E_{\text{inh}}$ and $E_{\text{m}}$, respectively. The effective bulk modulus of the composite is plotted for various values of $E_{\text{inh}}$. The effective bulk moduli obtained using numerical calculations, dilute approximation, using Eshelby's approach (Christensen, 2005), and the presented spectral

method (Eq. 14) are plotted for inhomogeneity over composite radii of (a) 0.2, (b) 0.5, and (c) 0.9. The values of the effective bulk moduli are normalized by the bulk modulus of the corresponding homogeneous solid, where $E_{\text{inh}} = E_{\text{m}}$.

## 2.4. Application to composites

We used Eq. (10) along with a series expansion to show that random composites become softer and less thermally conductive with increasing heterogeneity. In our analysis composites were made of a large number of inhomogeneities.

### 2.4.1. The weak contrast series expansion

We studied composites made by the coalescence of $N_S$ homogenous constituents. We denote the properties (elastic modulus or thermal conductivity) of the $i$'th constituent by $A_i$. The $A_i$ values can be considered as components of a vector, $x$, which describes the microstructural properties of the composite. We take the case of a homogenous material where all $A_i$'s are the same, as a reference. We refer to the effective properties and the properties of the constituents of the homogeneous material by $A_{\text{eff}}^{(0)}$ and $x_0$, respectively. In a generic composite, we express the effective properties and microstructural properties by $A_{\text{eff}}$ and $x_1$, respectively. We can express the effective properties of a composite with respect to the homogeneous sample using a series expansion as

$$A_{\text{eff}}(x_1) = A_{\text{eff}}^{(0)} + \left[(\delta x_1 \cdot \nabla_x) A_{\text{eff}} + \frac{1}{2}(\delta x_1 \cdot \nabla_x)^2 A_{\text{eff}}\right]_{x=x_0} + \cdots, \quad (15)$$

where $\delta x_1 = x_1 - x_0$. Similar weak contrast series expansions have been used in studying composite systems (Milton, 2002) and random networks of resistors (Luck, 1991).

In a random composite, multiple realizations of the composite and hence $x_1$, produce a distribution of $A_\text{eff}$ values. The variance of $A_\text{eff}$ can be found by calculating the variance of the series truncating the terms beyond the first order in $\delta x_1$, as $\sigma^2_{A_\text{eff}} \approx \sigma^2_{A_i}[\nabla_x A_\text{eff}]^2_{x=x_0}$. Here, $\sigma^2_{A_i}$ is the variance of the microstructural properties, which grows with increasing composite heterogeneity. Similarly, averaging the series truncated beyond the second order terms in $\delta x_1$, gives $A^{(0)}_\text{eff} - \langle A_\text{eff} \rangle \approx -(1/2)\sigma^2_{A_i}[\nabla^2_x A_\text{eff}]_{x=x_0}$, where $\langle A_\text{eff} \rangle$ denotes the averaged effective properties over an ensemble of composites with the same geometry, but with different realizations of microstructural properties, $x_1$. Equation (10)(10) indicates that $\nabla^2_x A_\text{eff}$ is negative in elastic composites. Therefore, $\langle A_\text{eff} \rangle < A^{(0)}_\text{eff}$, and the effective elastic stiffness of the composites decrease by increasing heterogeneity. Intuitively, the concave characteristic of the effective stiffness indicates that if the composite is comprised of two types of materials, with increasing the variance of the microstructural properties, the rate of gain in effective elastic stiffness from the stiffer constituents is smaller than the rate of its loss from the softer constituents.

**3. Comparison with numerical results**

This section presents a comparison of the theoretical results with numerical calculations. First, the numerical methods are presented. Next, the illustrative example of the effective bulk modulus of a composite containing an individual spherical inhomogeneity is described. Finally, two random composites are investigated: composites made by the coalescence of cubic blocks, and composites consisting of spherical inclusions in a homogenous matrix.

## 3.1. Numerical calculation method

The composite problems can be expressed as partial differential equations with spatially varying constants. The partial differential equations of elasticity and Fourier heat conduction were solved using the Abaqus (Hibbsett et al., 1998) and COMSOL Multiphysics software packages (COMSOL, 2014). Geometric models and unstructured meshes were generated using Abaqus and COMSOL. The problem domains were discretized into tetrahedral and hexahedral elements. Small enough elements were used to eliminate the mesh sensitivity of the reported results. Small-strain elasticity and steady state Fourier heat conduction were modeled by prescribing small displacement and temperature differences at the boundaries of the samples. The resulting forces and heat fluxes were integrated at the boundaries of the composites and were used for the evaluation of the effective properties. Implicit solution methods were used, and geometric nonlinearities were not considered.

## 3.2. The illustrative example of a spherical inhomogeneity in a volumetrically expanding composite

Numerical calculations were performed to verify the modal relation for the bulk modulus of a composite containing a spherical inhomogeneity. A geometric model was made comprising of two concentric spheres: the inner sphere represented the inhomogeneity, and the outer hollow sphere represented the surrounding matrix. The inhomogeneity and the surrounding matrix had elastic moduli $E_{\text{inh}}$ and $E_{\text{m}}$, radii $r_{\text{i}}$ and $r_{\text{o}}$, and Poisson's ratio 0.3. The $E_{\text{inh}}/E_{\text{m}}$ ratios $10^{-12}$, 5 and few values between them, and $r_{\text{i}}/r_{\text{o}}$ values of 0.2, 0.5, and 0.9 were tested. In every test, a small outward radial displacement, $d_0$, was prescribed to the outer surface of the composite, and the resulting traction was evaluated in the outer surface of the sample. A radial strain of 0.5% was prescribed in all cases. The effective bulk modulus was evaluated as the ratio of the mean surface traction to the volumetric strain. Small-strain deformation was considered, and the volumetric strain was evaluated as $3d_0/r_\mathrm{o}$.

The effective bulk modulus increased with increasing the stiffness of the inhomogeneity and exhibited a concave curve (Fig. 3). The same trend was observed when testing inhomogeneities of various radii. As expected, the change in effective bulk modulus was larger in composites with larger inhomogeneities. These results are compared with the theoretical results in the Discussion Section.

**3.3. Random composites made by the coalescence of cubic blocks**

To support the theoretical results on elastic random composites (Section 2.4) and empirically extend them to thermally conductive composites, numerical calculations were performed. Cubic representative volume elements were made by the coalescence of cubic homogeneous constituents of equal size. To demonstrate the effect of heterogeneity on the mean composite properties, the material properties of the constituents were sampled from statistical distributions of mean $A_\mathrm{eff}^{(0)}$ and variance, $\sigma_A^2$, where $\sigma_A^2$ was varied in calculations. For each value of $\sigma_A^2$, an ensemble of five hundred replicas of the composite were tested. Effective properties were measured in each of the five hundred samples. The mean and variance of the resulting distributions of effective properties, $\langle A_\mathrm{eff} \rangle$ and $\sigma_{A_\mathrm{eff}}^2$, were computed in terms of the variance of local properties, $\sigma_A^2$. The effective thermal conductivity was measured by the application of a small temperature difference, $\Delta T^{(0)}$, to two opposing faces of the composite. Insulating boundary conditions were prescribed at the other sample boundaries. The effective thermal conductivity, $k_{\text{eff}}$, was evaluated as $-QL/\Delta T^{(0)}$ where $Q$ is the heat flux per unit area over one of the surfaces of the composite with prescribed temperature.

The numerical calculations indicated that for an arbitrary $A_{\text{eff}}^{(0)}$, the mean of the distribution of composite properties decreases from the reference $A_{\text{eff}}^{(0)}$ in proportion to the variance of the microstructural properties: $A_{\text{eff}}^{(0)} - \langle A_{\text{eff}} \rangle \sim \sigma_A^2$ (Fig. 4a). In addition, the variance of effective properties increases linearly with the variance of microstructural properties: $\sigma_{A_{\text{eff}}}^2 \sim \sigma_A^2$ (Fig. 4b). Figure 4 includes the results for both elastic stiffness and thermal conductivity and for bimodal and lognormal distributions of microstructural properties. We observed that increasing the number of homogeneous constituents in the model leaves the results on composites unchanged. The data sets are described in the figure caption.

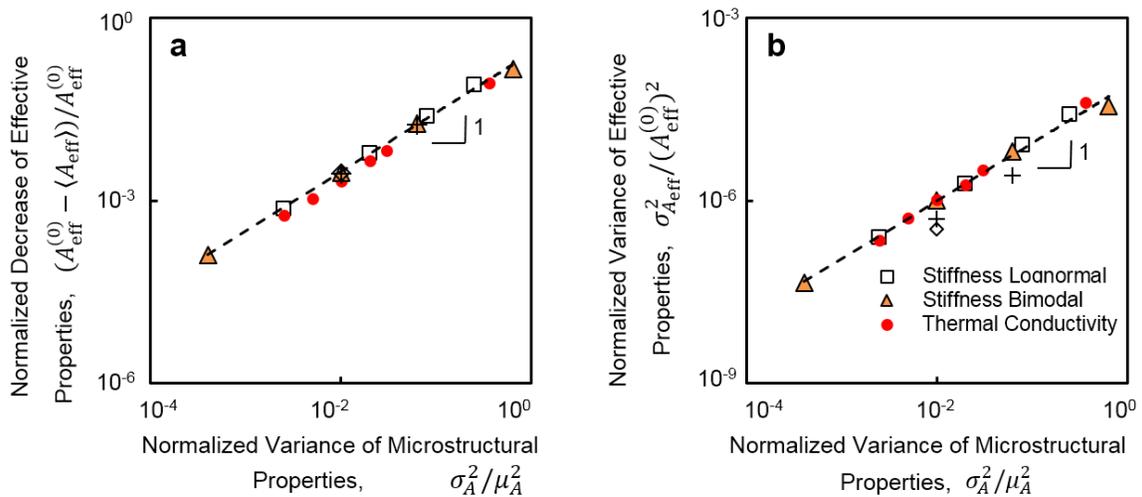

**Fig. 4.** (Color online) Application of the inhomogeneity relation to random composites made by the coalescence of cubic blocks shows that effective elastic stiffness and thermal conductivity

decrease with increasing heterogeneity in microstructural properties. Numerical results show (a) the reduction of the mean effective properties $\langle A_{\text{eff}} \rangle$ and (b) increase of the variance of effective properties $\sigma^2_{A_{\text{eff}}}$ with the variance of the microstructural properties, $\sigma^2_A/\mu^2_A$. The symbols represent data corresponding to the effective elastic stiffness of composites with lognormal (outlined squares) and bimodal (orange triangles) distributions of microstructural properties, and the effective thermal conductivity of composites with bimodal distributions of local thermal conductivity (red circles). Effective properties are plotted for composites that are twice (plus signs) and three times (open diamonds) larger than the original composite. In composites with bimodal distributions of properties, two types of constituents are present that have equal probabilities of presence. The data points and dashed lines correspond to the numerical results and analytical power laws, respectively.

We also tested the bulk modulus of the random composites made by the coalescence of cubic blocks (Fig. 5). The elastic moduli of the blocks were sampled from identical bimodal distributions of varying standard deviations. An ensemble of 50 samples with different realizations of microstructural properties were tested at each value of standard deviation. Mean effective bulk moduli were evaluated for each group of composites.

The mean bulk modulus of the composites decreased with increasing microstructural heterogeneity, and the amount of decrease was proportional to the variance of the microstructural properties (Fig. 5). The resulting mean effective bulk moduli resided between the upper and lower Voigt-Reuss bounds (Hill, 1963; Milton, 2002) and also the upper and lower Hashin-Shtrikman bounds (Hashin and Shtrikman, 1963) (Fig. 5).

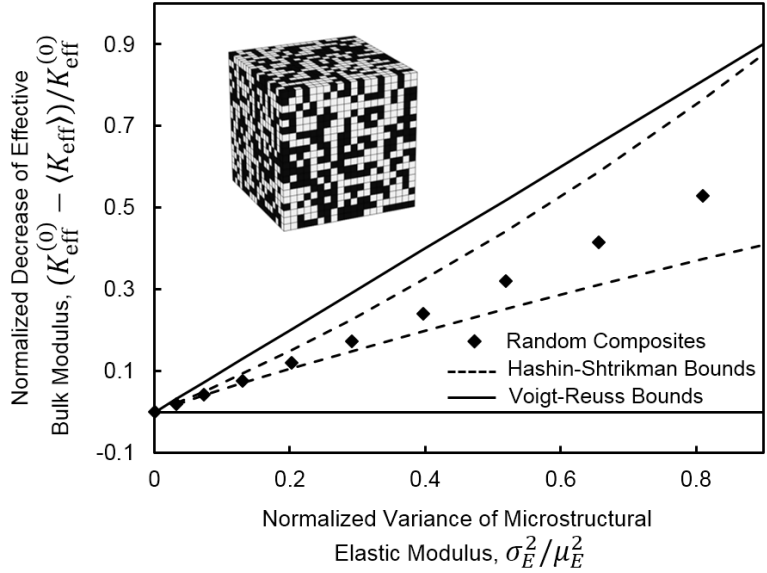

**Fig. 5.** Comparison of the mean effective bulk modulus of the random composites made by the coalescence of cubic blocks with the Hashin-Shtrikman and Voigt-Reuss bounds. The elastic moduli of the cubic blocks were sampled from identical bimodal distributions. The inset displays a snapshot of a tested representative volume element, where black and white colors represent constituents with two different elastic moduli.

Moreover, we tested random composites made from different numbers of constituents. The numerical results indicated that increasing the number constituents in the composite, leaves the decrease of effective properties unchanged (Fig. 4a). However, the variance of the effective properties is inversely related to the number of microstructural constituents, $N_S$ (Fig. 6a). This result is in agreement with the previous observations in elastic composites (Jeulin et al., 2004) and is a consequence of the central limit theorem (Dekking et al., 2007). Since the variance of effective properties becomes smaller as composites become larger, we conclude that in composites made of a large number of constituents, the mean of the effective properties is representative of the effective properties of all of the random composites in the ensemble.

In the absence of heterogeneity, strain is spatially uniform over the composite. In contrast, in random composites, the distribution of strain is nonuniform, and the variance of the strain field grows with increasing the heterogeneity of the composites (Fig. 6b). We calculated the variance of the strain field using the metric $\chi = \langle u_x'^2 \rangle_V / \varepsilon_0^2$ where $u_x' = u - u_{\text{uniform}}$ is the non-uniform component of displacement in the direction of the uniaxially applied macroscopic strain. $u$ and $u_{\text{uniform}}$ are the total and uniform components of the displacement. Here, $\langle u_x'^2 \rangle_V$ denotes the mean of $u_x'^2$ over the composite's volume. In conclusion, this result indicates that the growth of heterogeneity in the displacement field with microstructural heterogeneity agrees with the reported softening effect marking the departure of effective properties from the Voigt bound as heterogeneity increases.

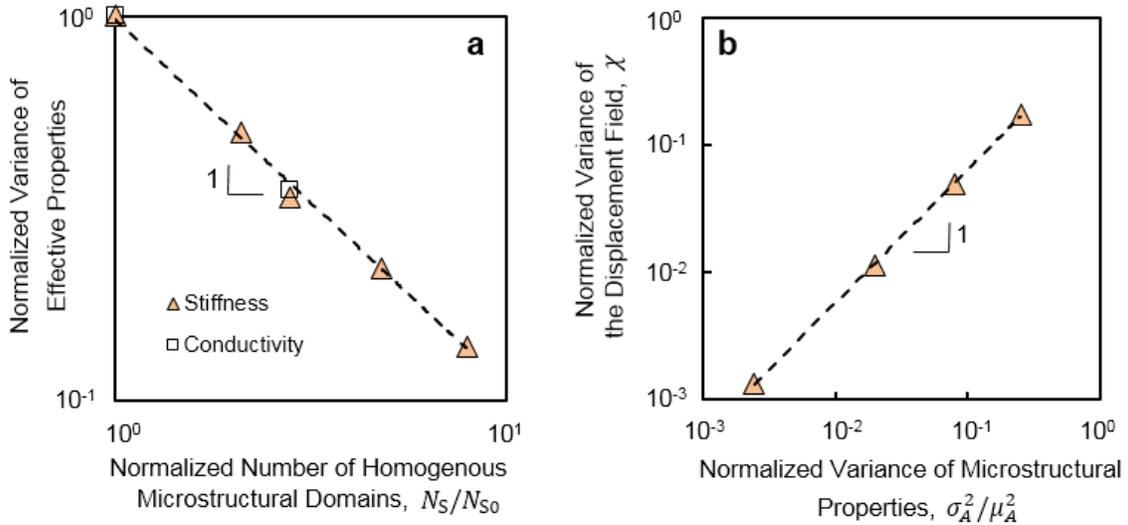

**Fig. 6.** (Color online) In the random composites made by the coalescence of cubic blocks, the variance of effective properties is inversely proportional to the number of homogeneous microstructural constituents, and the heterogeneity in the displacement field grows in proportion to the variance of the microstructural properties. The increase in the variance of displacement marks a departure from a uniform state of strain with increasing heterogeneity. All quantities were normalized by the respective values corresponding to the results of Fig. 4. The data points and dashed lines correspond to the numerical results and analytical power laws, respectively.

## 3.4. Random composites with spherical inclusions of random stiffnesses and thermal conductivities

We also tested the effective elastic stiffness and thermal conductivity of random composites consisting of spherical inclusions randomly placed in homogenous matrices. Spheres of uniform volume and random properties were placed in a cubic representative volume element by the random choice of the Cartesian coordinates of their centers. In different sets of tests, inclusions were modeled that each occupied 0.002 or 0.003 of the composite's volume. The elastic moduli or thermal conductivities of the inclusions were independently sampled from identical bimodal distributions. Both the elastic inclusions and the surrounding matrix had Poisson's ratio of 0.3. The mean elastic modulus or thermal conductivity of the inclusions was 20 times that of the surrounding matrix. Tests were performed by prescribing a small displacement or temperature difference at the sample boundaries. The same boundary conditions used in the case of cubic composites were used here, and the effective properties were evaluated using the resulting boundary traction forces or heat fluxes. Ensembles of 50 composites were tested at each variance of inclusion properties, where different realizations of inclusion properties were present in each test. The matrix properties were the same in all tests.

The thermally conductive composites became less conductive on average with increasing the variance of the thermal conductivity of the inclusions. At small to moderate variances of the properties of the inclusions, the decrease in the effective thermal conductivity was proportional to the variance of the properties of the inclusions (Fig. 7).

Similarly, we tested elastic composites with inclusions that in total occupied 1% and 10% of the sample volume. In the case with 10% volume fraction, $V_i/V$, we tested composites with different numbers and sizes of individual inclusions. The first case had a smaller number of large inclusions, and the second case had more inclusions of a smaller size. In all cases, effective stiffness decreased with increasing the variance of the elastic moduli of the inclusions. Furthermore, at the small and moderate variances of the elastic modulus of the inclusions, the decrease in effective stiffness was proportional to the variance of the moduli of the inclusions (Fig. 7). The effective stiffness of the composites with a higher volume fraction of inclusions decreased more than the composites with a smaller volume fraction of inclusions. At the same total volume fraction, however, the relative decrease of effective stiffness was independent of the number and size of the individual inclusions. Taken together these tests demonstrate that the decrease in effective properties is also present in random composites with spherical inclusions of random properties.

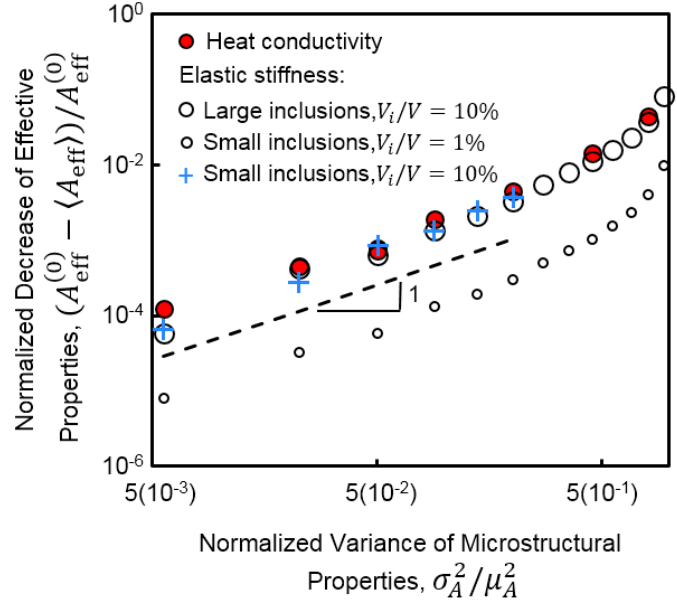

**Fig. 7.** (Color online) The effective elastic stiffness and heat conductivity of random composites with spherical inclusions with random properties decreases in proportion to the variance of the properties of the inclusions. The relative decrease of the effective elastic stiffness depends on the total volume fraction of inclusions, $V_i/V$. But at the same total volume fraction of inclusions, the relative decrease is independent of the number of inclusions and their size. The data points and dashed lines correspond to the numerical results and analytical power laws, respectively.

## 4. Discussion

To illustrate the modal approach and validate it, we evaluated the change in the effective bulk modulus of a composite containing an individual spherical inhomogeneity. The results showed a close agreement between the numerical calculations and the modal approach (Fig. 3). We also compared the result with a dilute approximation of effective bulk modulus based on the solution for spherical inhomogeneities (Fig. 3, red dashed curves), derived using Eshelby's approach (Christensen, 2005). As expected the dilute approximation agrees with the effective bulk moduli of the composites with small volume fractions of inhomogeneities. In the systems where the inhomogeneity radius is 90% of the composite radius, however, the dilute approximation predicts effective bulk moduli that are smaller than the bulk moduli obtained using numerical calculations and the modal relation (Fig. 3). The difference between the results increases with increasing the difference between the properties of the inhomogeneity and the surrounding matrix. In contrast, the numerical calculations and the modal relation closely agree even in the tests performed at a large volume fraction of the inhomogeneity (Fig. 3). This example demonstrates the validity of the modal derivation in finite-sized composites.

The studied composites have constituents whose properties are independently sampled from identical statistical distributions. Our results are, therefore, expected to apply to composites without long-range correlations of microstructural properties. It is expected that if long-range correlations are introduced to bias the sampling of the microstructural properties, the effective elastic stiffness still decreases with increasing heterogeneity. The averaging calculations, however, will no longer be valid, and we expect that the amount of decrease in stiffness will no longer be proportional to the variance of the microstructural properties. Moreover, in cases where a gradient of material properties is present, we expect a decrease of effective stiffness with increasing heterogeneity, at the same mean of properties, whereas we do not expect the amount of decrease to be proportional to the variance of the microstructural properties. Future works may extend the present results to heterogeneous media with spatially correlated microstructural properties.

The modal relation shows that the effective stiffness is a concave function of the elastic modulus of an inhomogeneity inside a composite (Eq. (10)). The derivation also points to an exceptional case where the second derivative of effective stiffness with respect to the elastic modulus of the inhomogeneity vanishes. A system of elastic rods (or Hookean springs) connected in parallel, is an example of this exceptional case, where the deformation of the inhomogeneity is prescribed by the boundary conditions, and all $c_j$'s vanish. Therefore, $\partial^2 C_{\text{eff}}/\partial E_{\text{inh}}^2 = 0$, and the effective stiffness of the system does not change with increasing the variance of the elastic modulus of the rods, at the same mean of constituent properties. In this exceptional system, after external displacement is prescribed to evaluate the effective properties, the individual constituents cannot be further deformed by the application of external forces that correspond to the modes that comprise the deformation of the constituent.

The developed modal approach presents insights into the mechanics of inhomogeneities and random composites. Its extension to different classes of composites, however, requires the determination of the invariants $c_j$ for specific composite geometries. The $c_j$ parameters are invariant of the inhomogeneity's modulus, whereas they depend on the shape of the inhomogeneity, the response function of the surrounding medium, and the prescribed boundary conditions. Further developments of the modal approach may lead to novel solutions for inhomogeneity problems and composites.

**5. Conclusions**

This paper presented a modal analysis of the influence of an inhomogeneity on the effective elastic stiffness of a composite. The analysis showed that effective stiffness is always a concave function of the elastic modulus of an inhomogeneity inside the composite. The concave property indicates that weakly heterogenous random composites become softer with increasing microstructural heterogeneity; the loss of the stiffness by the softer constituents is larger than the gain from the stiffer constituents. The modal relation was illustrated and validated in the example case of a spherical inhomogeneity in a volumetrically expanding composite. The results on random composites were validated using numerical calculations and were compared with the Voigt-Reuss and Hashin-Shtrikman bounds. Moreover, the numerical results showed that random composites including spherical inclusions of random properties become softer with increasing the variance of the elastic moduli of the inhomogeneities. Finally, the results were numerically extended to thermally conductive composites, indicating a decrease in effective thermal conductivity with increasing microstructural heterogeneity.

**Acknowledgments**
We thank Javad Heydari (formerly at RPI) and Poorya Mirkhosravi (UCSD) for fruitful discussions. We thank Professor Catalin R. Picu (RPI) for guidance.

**Funding**
This research did not receive any specific grant from funding agencies in the public, commercial, or not-for-profit sectors.

Bower, A.F., 2009. Applied Mechanics of Solids. CRC Press.
Castañeda, P.P., 1991. The effective mechanical properties of nonlinear isotropic composites. J. Mech. Phys. Solids 39, 45–71. https://doi.org/10.1016/0022-5096(91)90030-R
Christensen, R., 2005. Mechanics of Composite Materials. Dover Publications, Mineola, N.Y.
Chung, D., 2010. Composite Materials: Science and Applications, 2nd ed. 2010 edition. ed. Springer, London; New York.
Ciarlet, P.G., 2013. Linear and Nonlinear Functional Analysis with Applications. SIAM, Philadelphia, PA.
COMSOL Multiphysics Reference Manual, version 5, 2014.
Dekking, F.M., Kraaikamp, C., Lopuhaä, H.P., Meester, L.E., 2007. A Modern Introduction to Probability and Statistics: Understanding Why and How. Springer, London.
Dimas, L.S., Veneziano, D., Giesa, T., Buehler, M.J., 2015a. Probability distribution of fracture elongation, strength and toughness of notched rectangular blocks with lognormal Young's modulus. J. Mech. Phys. Solids 84, 116–129. https://doi.org/10.1016/j.jmps.2015.06.016
Dimas, L.S., Veneziano, D., Giesa, T., Buehler, M.J., 2015b. Random Bulk Properties of Heterogeneous Rectangular Blocks With Lognormal Young's Modulus: Effective Moduli. J. Appl. Mech. 82, 011003–011003. https://doi.org/10.1115/1.4028783
Eshelby, J.D., 1959. The elastic field outside an ellipsoidal inclusion. Proc R Soc Lond A 252, 561–569. https://doi.org/10.1098/rspa.1959.0173
Eshelby, J.D., 1957. The Determination of the Elastic Field of an Ellipsoidal Inclusion, and Related Problems. Proc. R. Soc. Lond. Math. Phys. Eng. Sci. 241, 376–396. https://doi.org/10.1098/rspa.1957.0133
Gubernatis, J.E., Krumhansl, J.A., 1975. Macroscopic engineering properties of polycrystalline materials: Elastic properties. J. Appl. Phys. 46, 1875–1883. https://doi.org/10.1063/1.321884
Hashin, Z., Shtrikman, S., 1963. A variational approach to the theory of the elastic behaviour of multiphase materials. J. Mech. Phys. Solids 11, 127–140. https://doi.org/10.1016/0022-5096(63)90060-7
Hibbett, Karlsson, Sorensen, 1998. ABAQUS/standard: User's Manual. Hibbitt, Karlsson & Sorensen.
Hill, R., 1963. Elastic properties of reinforced solids: Some theoretical principles. J. Mech. Phys. Solids 11, 357–372. https://doi.org/10.1016/0022-5096(63)90036-X
Jeulin, D., Kanit, T., Forest, S., 2004. Representative Volume Element: A Statistical Point of View, in: Bergman, D.J., Inan, E. (Eds.), Continuum Models and Discrete Systems, NATO Science Series. Springer Netherlands, pp. 21–27. https://doi.org/10.1007/978-1-4020-2316-3_5
Lebedev, L., Vorovich, I., n.d. Functional Analysis in Mechanics. Springer, New York.
Lopez-Pamies, O., Goudarzi, T., Nakamura, T., 2013. The nonlinear elastic response of suspensions of rigid inclusions in rubber: I—An exact result
35